\documentclass[aps,prl,twocolumn,showpacs,superscriptaddress]{revtex4}
\usepackage{graphicx}
\usepackage{bm}

\newcommand{\beq}{\begin{equation}}
\newcommand{\eeq}{\end{equation}}

\begin{document}

\title{\boldmath  Thermodynamic Properties of Fermi Systems with Flat
Single-Particle Spectra}
\author{V.~A.~Khodel}
\affiliation{ Russian Research Centre Kurchatov
Institute, Moscow, 123182, Russia}
\affiliation{ McDonnell Center for the Space Sciences and
Department of Physics, Washington University,
St.~Louis, MO 63130, USA }
\author{J.~W.~Clark}
\affiliation{ McDonnell Center for the Space Sciences and
Department of Physics, Washington University, St.~Louis, MO 63130,
USA }
\author{M.~V.~Zverev}
\affiliation{ Russian Research Centre Kurchatov
Institute, Moscow, 123182, Russia}

\date{\today}

\begin{abstract}
The behavior of strongly correlated Fermi systems is investigated
beyond the onset of a phase transition where the single-particle
spectrum $\xi({\bf p})$ becomes flat.  The Landau-Migdal quasiparticle
picture is shown to remain applicable on the ordered side of this transition.
Nevertheless, low-temperature properties
evaluated within this picture show profound changes relative
to results of Landau theory, as a direct consequence of the flattening
of $\xi({\bf p})$.  Stability conditions for this class of systems
are examined, and the nature
of antiferromagnetic quantum phase transitions is elucidated.

\end{abstract}

\pacs{
71.10.Hf,
71.27.+a
}
\maketitle

Low-temperature properties of strongly correlated Fermi systems
exhibit features not inherent in Landau Fermi liquids
\cite{godfrin,morishita,saunders,loh,coleman1,stewart,steglich,takahashi,gegenwart,custers}.
Of special importance is the scaling behavior
$\chi^{-1}(T,H)=\chi^{-1}(0,H)+T^{\alpha}F(H/T)$ of the inverse
magnetic susceptibility, observed so far only in  heavy fermion metals
\cite{coleman1,steglich,takahashi,gegenwart,custers} in weak,
sometimes tiny external magnetic fields $H$.  This feature rules
out the collective, spin-fluctuation scenario advanced in
Refs.~\onlinecite{hertz,millis} to explain the non-Fermi-liquid
(NFL) behavior of these systems, while providing evidence for the
direct relevance of single-particle degrees.
Earlier work within this picture \cite{ckz} has
evaluated thermodynamic properties on the ``metallic'' side of a
phase transition associated with the rearrangement of the Landau
state in strongly correlated Fermi systems. Here we focus
attention on the situation beyond the critical point where the
Landau state becomes unstable.

This state is known (e.g.\ from Refs.~\onlinecite{zb,baldo})
to lose its stability at a density $n=n_b$ where a bifurcation point
$p=p_b$ emerges in the equation
\beq
\xi(p,n,T=0)=0\ ,
\label{root}
\eeq
which ordinarily has only the single root
$p=p_F$. (Here $\xi(p)=\varepsilon(p)-\mu$ is the single-particle
energy measured from the chemical potential $\mu$.)  We shall
focus on the case $|p_F-p_b|\ll p_F$, where the critical
single-particle spectrum has the form
\beq
\xi(p,n_b,T=0)\sim(p-p_b)^2(p-p_F) \ .
\label{spnb}
\eeq
Thus if $p_b$ coincides with $p_F$, the effective mass $M^*(n)$ diverges
at $n_b$; in the general case $p_b\neq p_F$, the Landau state
loses its stability before $M^*$ becomes infinite. Beyond the
critical density $n_b$, Eq.~(\ref{root}) possesses two additional
roots $p_1$ and $p_2$ with $p_1<p_b<p_2$. The spectrum
$\xi_{FL}(p,T=0)$, evaluated with the FL momentum distribution
$n_{FL}(p)=\theta(p_F-p)$, then takes the form
\beq
\xi_{FL}(p,n,T=0)\sim (p-p_1)(p-p_2)(p-p_F) \ .
\label{root1}
\eeq
If $p_b\neq p_F$, the roots $p_1$, $p_2$ are both located either in
the interior of the Fermi sphere or both outside it.  If
$p_b=p_F$, then $p_1<p_F<p_2$.  In all these cases, the occupation
numbers $n_{FL}(p)$ are rearranged. As a rule, the Fermi surface
becomes multi-connected, but at $T=0$, the quasiparticle
occupation numbers  continue to take values 0 or 1. Hence the
Landau-Migdal quasiparticle picture holds, with $n(\xi)=1$ for
$\xi<0$ and 0 otherwise.  Consider first the case $p_1<p_2<p_F$.
Then according to Eq.~(\ref{root1}), the single-particle states
remain filled in the intervals $p<p_1$ and $p_2<p<p_F$, while the
states with $p_1<p<p_2$ are empty. We call this new phase the
bubble phase. If $p_b=p_F$, then $p_1{<}p_F$ and $p_2{>}p_F$, and
the states with $p{<}p_1$ and with $p_F{<}p{<}p_2$ are occupied,
while those for $p_1{<}p{<}p_F$ are empty.  Again one deals with
the bubble phase.

\begin{figure}[t]
\includegraphics[width=0.7\linewidth,height=1.0\linewidth]{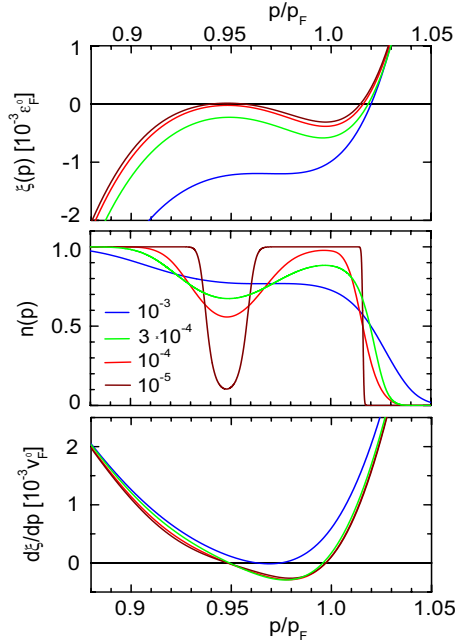}
\caption{Single-particle spectrum $\xi(p)$ in units of
$10^{-3}\,\varepsilon_F^0$, where $\varepsilon_F^0=p^2_F/2M$ (top
panel), occupation numbers $n(p)$ (middle panel), and $d\xi/dp$ in
units of $10^{-3}\,v_F^0$, where $v_F^0=p_F/M$ (bottom panel),
plotted versus $p/p_F$ at four color-coded temperatures relevant
to the bubble phase, in units of $\varepsilon_F^0$. The model
(\ref{mod0}) is assumed with parameters $\beta_1=0.07$ and
$\lambda_1=0.45\,N_0$, where $N_0=p_FM/\pi^2$.} \label{fig:0pfb}
\end{figure}

\begin{figure}[t]
\includegraphics[width=0.7\linewidth,height=1.0\linewidth]{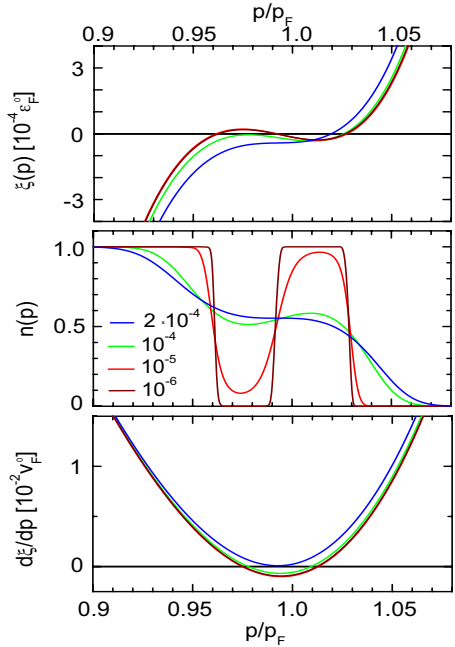}
\caption{Same as in Fig.~\ref{fig:0pfb} but for the model
(\ref{model}) with parameters $\lambda_2=3\,N_0$ and
$\beta_2=0.48$. } \label{fig:2pfb}
\end{figure}

\begin{figure}[t]
\includegraphics[width=0.95\linewidth,height=0.9\linewidth]{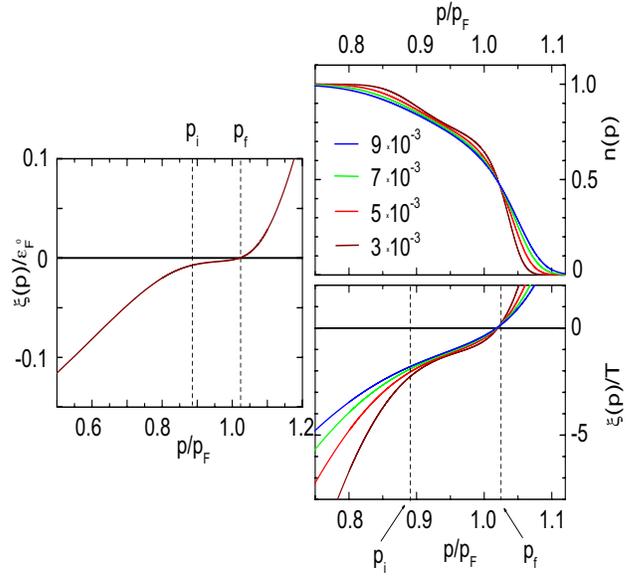}
\caption{Single-particle spectrum $\xi(p)$ in units of
$\varepsilon_F^0$
at the critical temperature
$T_Z=3\times 10^{-3}\,\varepsilon^0_F$ (left panel), occupation
numbers $n(p)$ (right-top panel), and $\xi(p)/T$ (right-bottom
panel), plotted versus $p/p_F$ at four color-coded temperatures
relevant to the phase with a fermion condensate, in units of
$\varepsilon_F^0$. The model (\ref{mod0}) is assumed.}
\label{fig:0pffc}
\end{figure}

\begin{figure}[t]
\includegraphics[width=0.95\linewidth,height=0.9\linewidth]{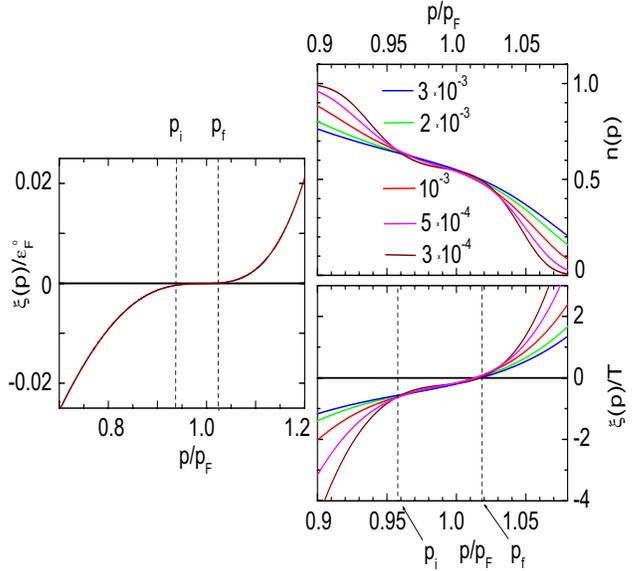}
\caption{Same as in Fig.~\ref{fig:0pffc} but for the model
(\ref{model}). The single-particle spectrum in the left panel is
shown at $T_Z=3\times 10^{-4}\,\varepsilon^0_F$. }
\label{fig:2pffc}
\end{figure}

At this point, we observe that the solution (\ref{root1}) is not
self-consistent, since the spectrum is evaluated with $n_{FL}(p)$
while the true Fermi surface is doubly-connected. Following
Ref.~\onlinecite{zb}, we consider the feedback of the
rearrangement of $n_{FL}(p)$ on the spectrum $\xi(p)$ in the
bubble phase based on the Landau relation \cite{lanl,trio}
\beq
{\partial\varepsilon(p)\over\partial {\bf p}}={{\bf p}\over
M} +\int f({\bf p},{\bf p}_1) {\partial n(p_1)\over
\partial {\bf p}_1}d\upsilon_1 \ ,
\label{lansp}
\eeq
where $f$ is the scalar
part of the Landau interaction function and
$n(p)=[1+\exp(\xi/T)]^{-1}$ is the quasiparticle momentum
distribution.
For the electron liquid within a solid, ${\bf p}/M$ is to
be replaced by $\partial\varepsilon^0_{{\bf p}}/\partial {\bf p}$,
where $\varepsilon^0_{{\bf p}}$ is the LDA electron spectrum.
To date, Eq.~(\ref{lansp}) has been solved only in 3D Fermi systems
with functions $f$ depending on $q=|{\bf p}-{\bf p}_1|$.
Despite the diversity of forms assumed for $f(q)$,
the resulting spectra and momentum distributions
bear a close family resemblance.   This robustness is illustrated
in Figs.~\ref{fig:0pfb} and \ref{fig:0pffc}, which display results
\cite{zb} from solution of Eq.~(\ref{lansp}) for the interaction
function
\beq f(q)=\lambda_1/[(q/p_F)^2+\beta_1^2] \ ,
\label{mod0}
\eeq
and in Figs.~\ref{fig:2pfb} and \ref{fig:2pffc},
which present results for
\beq
f(q)=\lambda_2/[((q/2p_F)^2-1)^2+\beta_2^2]\ .
\label{model}
\eeq

How do the bubble solutions of Eq.~(\ref{lansp}) evolve under
variation of $T$?  At extremely low $T<T_{FL}\sim (p_2-p_1)^2/M$,
these states are described by FL theory with the enhanced
effective mass $M^*\sim Mp_F/(p_2-p_1)$. Heating above $T_{FL}$
results in their dissolution (see Figs.~\ref{fig:0pfb} and
\ref{fig:2pfb}). With further increase of $T$, the spectrum
$\xi(p)$ becomes smoother, and in the region of a new critical
temperature $T_Z$, a flat portion $\xi\simeq 0$ appears over an
interval $[p_i,p_f]$ surrounding the point $p_F$, as shown in the
left panels of Figs.~\ref{fig:0pffc} and \ref{fig:2pffc}. The
presence of this flat portion of $\xi(p)$ is a signature of the
phenomenon called fermion condensation \cite{ks,vol,noz}. Since
$\xi(p)=\varepsilon(p)-\mu$ and $\varepsilon(p)=\delta E_0/\delta
n(p)$, the equality $\xi=0$ can be rewritten as a variational
condition
\beq {\delta E_0\over \delta n(p)}=\mu \ , \quad
p_i<p<p_f \ ,
\label{var}
\eeq
with $E_0=\sum_{{\bf p}}\varepsilon^0_{{\bf p}}n({\bf p}) +
{1\over 2} \sum_{{\bf p},{\bf p}_1}f({\bf p}-{\bf p}_1) n({\bf p})
n({\bf p}_1)$ and
$\varepsilon^0_{\bf p} = p^2/2M$.  The solution $n_0(p)$ of
Eq.~(\ref{lansp}), or equivalently of Eq.~(\ref{var}), is a
continuous function of $p$ with a nonzero derivative $dn_0/dp$.
The set of  states with $\xi(p)=0$ is called the fermion
condensate (FC), since the associated density of states contains a
Bose-liquid-like term $\eta n\delta(\varepsilon)$. The
dimensionless constant $\eta\simeq (p_f-p_i)/p_F$ is identified as
a characteristic parameter of the FC phase.

It has been shown \cite{noz} that the FC ``plateau'' in $\xi(p)$
has a small slope, evaluated by inserting $n_0(p)$ into the above
Fermi-Dirac formula for $n(\xi)$ to yield
\beq \xi(p,T\geq T_Z)=T\ln {1-n_0(p)\over n_0(p)} \,, \,\,
p_i<p<p_f \, .
\label{spt}
\eeq As
indicated in the bottom-right panels of
Figs.~\ref{fig:0pffc} and \ref{fig:2pffc}, at  $T\geq T_Z$ the
ratio $\xi(p)/T$ is indeed a $T$-independent function of $p$ in
the FC region. The width $\xi(p_f)-\xi(p_i)\equiv \xi_f-\xi_i$ of
the FC ``band''  appears to be of order $T$, almost independently
of $\eta> \eta_{\mbox{\scriptsize min}}\sim 10^{-2}$. Thus at
$\eta>\eta_{\mbox{\scriptsize min}}$ the
FC group velocity is estimated as
\beq
\left({d\xi(p,T)\over dp}\right)_T\sim {T\over \eta p_F} \ ,
\quad p_i<p<p_f \ .
\label{estfc}
\eeq

Outside the FC domain, $\xi(p)$ still remains flat (see
Figs.~3, 4), and contributions to thermodynamic properties from the
regions adjacent to $ p_i$ and $ p_f$ also play significant roles
in their NFL behavior.  Indeed, consider for example the static
magnetic susceptibility \beq \chi(T)=-\mu^2_B\Pi_0(\omega= 0) [
1- g_0 \Pi_0(\omega=0)]^{-1} \,. \label{chio} \eeq Here $\mu_B$
denotes the Bohr magneton \cite{magneton} and $g_0$ the zeroth harmonic of the
spin-spin interaction function, while $\Pi_0(\omega= 0)=\int
\left(dn(\xi)/d\xi\right) d\upsilon$ is the static particle-hole
propagator.

The propagator $\Pi_0(\omega=0)\equiv -N_0(P_f(T)+P_n(T))/T$,
where $N_0$ is the density of states of the ideal Fermi gas, is
the sum of a FC part given by
\beq P_f(T)= {1\over p_F} \int\limits_{p_i}^{p_f}
n_0(p)\,[1-n_0(p)]\, dp\sim \eta
\label{pic}
\eeq
and a noncondensate part $P_n(T)\equiv P_n^>(T)+P_n^<(T)$
consisting of two terms.  Defining
\beq
P_n(T;\xi_1,\xi_2) = {1\over p_F}\int\limits_{\xi_1}^{\xi_2}
{n(\xi)\,[1-n(\xi)]\,d\xi \over \left(d\xi/ dp\right)}\  , \label{pin}
\eeq the two terms become $P_n^< =  P_n(T;-\mu,\xi_i)$ and
$P_n^>=P_n(T;\xi_f,\infty)$.  According to Eq.~(\ref{spnb}), at
small $\eta$ one has
\beq
d\xi(p\to p_f)/dp=v_f(T)+v_2(p-p_f)^{s-1}+ \dots \ ,
\label{groupn}
\eeq
with $v_f(T)\sim T$ (see Eq.~(\ref{estfc})) and $s=2$ and 3,
respectively, for the cases $p_b\neq p_F$ and $p_b=p_F$.   An
analogous formula applies for the group velocity $d\xi/dp$ outside
the FC domain close to the point $p_i$.  Based on these results,
algebra similar to that performed in Ref.~\onlinecite{ckz} leads to
\beq
\Pi_0(\omega=0)=-N_0\tau^{-1}[P_f\eta+P_n\tau^{1/s}]+{\rm const} \,,
\label{sf}
\eeq
where $\tau=T/\varepsilon^0_F$ is the dimensionless temperature.

The NFL excess $\Delta\chi(T,\rho)=\chi(T,\rho)-\chi_{FL}(\rho)$
over the Pauli result then acquires the form
\beq
\Delta\chi(T,\rho)\sim\mu_B^2N_0\tau^{-1}
{P_f\eta+P_n\tau^{1/s}\over
1{+}g_0(\rho)N_0\tau^{-1}(P_f\eta{+}P_n\tau^{1/s})} \,.
\label{dchi}
\eeq
We see that at small $\eta<\tau^{1/s}$ the FC plays a minor
role, and the NFL part of $\chi$ behaves as \beq\chi(T)\sim
T^{-1+1/s} \  . \label{nfl1} \eeq By contrast, in the case
$\eta>\tau^{1/s}$ the FC contribution to Eq.~(\ref{dchi}) is
predominant, and the magnetic susceptibility mimics that of a gas of
localized spins
\beq
\chi(T)\sim ( T-\Theta_W)^{-1} \,,
\label{cw}
\eeq with the Weiss temperature $\Theta_W\simeq-g_0\,\eta(\rho)\,\rho$.

\begin{figure}[t]
\includegraphics[width=0.7\linewidth,height=0.5\linewidth]{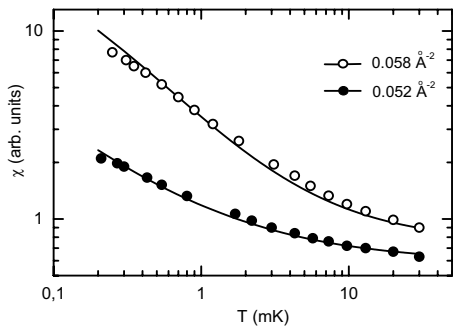}
\caption{Low-temperature magnetic susceptibility of $^3$He films.
Experimental data from Ref.~\onlinecite{godfrin} appear as solid
and open circles and solid squares, while solid curves trace the
predictions of the current theory at low $T$. }
\label{fig:chi2}
\end{figure}

In heavy-fermion metals, both index regimes, i.e.\ $s=2$
\cite{loh} and $s=3$ \cite{steglich}, are present, while data on
the magnetic susceptibility of 2D liquid $^3$He are compatible only with
$s=3$.  Fig.~\ref{fig:chi2} compares results from
Eq.~(\ref{dchi}) for 2D liquid $^3$He with
experimental data of Ref.~\onlinecite{godfrin} at the densities
$\rho=0.052$\,\AA~and $\rho=0.058$\,\AA.  We have made these
parameter choices: $P_n =0.2$, $P_f=1$, $\eta=0$ (lower curve),
and $\eta=0.04$ (upper curve).  The theoretical results are
seen to be in agreement with the experimental data.

The FC contributions to other thermodynamic properties are found
by inserting the distribution $n_0(p)$ into the corresponding
Landau formulas.  In particular, the FC entropy $S_f$ arising at
$T\simeq T_Z$ is  $S_f=-\sum \left [n_0(p)\ln n_0(p)+(1-n_0(p))\ln
(1-n_0(p) \right ]\sim \eta$.  This term does not contribute to
the specific heat $C(T)=T\partial S(T,\rho)/\partial T$; however,
it does affect the thermal expansion $\beta(T)\sim\partial
S(T,\rho)/\partial \rho$ \cite{zksb}, giving rise to a great
enhancement of the Gr\"uneisen ratio $\beta(T)/C(T)$, observed at
low $T$  in several heavy-fermion metals
\cite{steglich,steglich2}.

Our analysis has been carried out within the Landau-Migdal
quasiparticle picture where damping $\gamma$ of single-particle
excitations is neglected.  It is applicable to systems containing
a FC, provided the dimensionless damping rate $\gamma(T)/T$
remains small at $T\sim T_Z$.  To establish this property, we
adopt the standard formula \cite{trio} \beq
\gamma(\varepsilon{\sim}T )\sim \sum_{\bf
q}\!\!\int\limits_0^{\varepsilon\sim T}\!\!\!\!
|\Gamma^2(q,\omega)| {\rm Im} \Pi_0({\bf q},\omega) {\rm Im}
G_R({\bf p}-{\bf q},\omega-\varepsilon) d\omega \   ,
\label{imsig} \eeq  $\Gamma$ being the scattering amplitude and
$G_R(p,\varepsilon)$ the retarded Green function.   With a FC
present, overwhelming contributions to the collision term
(\ref{imsig}) are known to come from a region $q<q_c=\eta p_F$
where ${\rm Im} \Pi_0({\bf q},\omega)$ substantially exceeds the
ideal-Fermi-gas value \cite{noz}.  Factoring out $|\Gamma|^2$ from
the integral (\ref{imsig}), a straightforward argument \cite{noz}
yields $\gamma(T)\sim T^{-1}$.  However, this treatment is
erroneous, since it misses the suppression of the integral
(\ref{imsig}) stemming from the inequality $|\Gamma(q,\omega)|<
\left[{\rm Im}\Pi_0(q,\omega)\right]^{-1}$ \cite{schuck}. Upon
inserting this inequality into Eq.~(\ref{imsig}) we are led to the
result
\beq
\gamma( T) <\int\limits_{q<q_c}\int\limits_0^T
{ {\rm Im}\,G_R({\bf p} {+}{\bf q}, \varepsilon{+}\omega)\over
|{\rm Im}\,\Pi_0(q,\omega\sim T)|}\, d{\bf q}\,d\omega \ .
\label{imsig2}
\eeq
In the 2D electron gas this formula coincides with that
derived in Ref.~\cite{quinn}.  Based on Eq.~(\ref{imsig2}), the
putative behavior $\gamma(T) \sim T^{-1}$ is replaced by
\beq \gamma(T)\sim T\,\eta\,\ln(1/\eta) \  . \label{bcrfc} \eeq In
deriving this result, we have taken account of the fact that in
the system with a FC, the Fermi velocity $v_F=p_F/M^*$ entering
the corresponding expression for $\gamma(T)$ in Ref.~\cite{quinn}
must be replaced by the FC group velocity $(d\xi/dp)_T=T/\eta
p_F$. (More details may be found in Ref.~\cite {kcz}.)  Similar
results are obtained in the 3D case, but without the logarithmic
correction $\ln (1/\eta)$.  We conclude that the quasiparticle
picture holds in systems with a FC, at least when $\eta\ll 1$.

Finally, we consider stability conditions for systems with a
flat portion in the spectrum $\xi(p)$.  To be definite, we examine
the spin-density wave channel and suppose that at the critical
density $n_b$, the associated stability condition, which has the form
\beq
1>g_0(k_c)\Pi_0(k_c,\omega{=}0)\equiv g_0(k_c)\!
\int {n({\bf p}){-}n({\bf p}{+}{\bf k_c})\over
\xi({\bf p}){-}\xi({\bf p}{+}{\bf k_c})}\,d\upsilon \ ,
\label{sc}
\eeq
is not yet violated.  Beyond the density $n_b$, the magnitude of the
NFL component of $\Pi_0$, proportional to $\eta$, drops rapidly with
increasing $T$ (see e.g.~Eq.~(\ref{sf})).
In the present case this implies that violation of the inequality
(\ref{sc}) can occur only at very low $T$.  Furthermore, numerical
calculations show that in systems with a FC, the NFL part of the
function $\Pi_0(k,\omega=0)$ has a maximum at small nonzero $k\leq \eta
p_F$; one expects a new ground state possessing a long-range magnetic
superstructure, as well as a small ordered magnetic moment.
A salient feature revealed in Refs.~\cite{shag,zk,ckz} is
the destruction of the flat portion of the
spectrum $\xi(p)$ by imposition of sufficiently weak magnetic
fields, which also kills the magnetic ordering.  Such a quantum
phase transition has been uncovered in recent experiments
on YbAgGe \cite{fak}. The
interplay between quantum antiferromagnetism built upon a FC, and
the FC itself, will be the subject of a future article.

The present investigation has revealed generic features of
strongly interacting Fermi systems which exhibit a flat
single-particle spectrum $\xi(p)$.  We have shown that
the quasiparticle picture
remains applicable in the evaluation of
thermodynamic properties of such systems, and that the flattening
of $\xi(p)$ is a principal source of their non-Fermi-liquid
behavior.

We thank V.~V.~Borisov, P.~Coleman, V.~M.~Galitski, V.~G.~Orlov,
and V.~M.~Yakovenko for valuable discussions. This research was
supported by NSF Grant PHY-0140316 (JWC and VAK), by the McDonnell
Center for the Space Sciences (VAK), and by the Grant
NS-1885.2003.2 from the Russian Ministry of Education and Science
(VAK and MVZ).


\begin{thebibliography}{99}
\bibitem{magneton} In heavy-fermion metals, plane waves are replaced by
Bloch waves and $\mu_B$ by $\mu_{\mbox{\scriptsize eff}}$.
\end{thebibliography}
\end{document}